\documentclass[journal]{IEEEtran}
\usepackage{graphicx}
\usepackage{subfigure}
\usepackage{fancyhdr}
\usepackage{makeidx}
\usepackage{amssymb,amsmath}
\usepackage{balance}
\usepackage{color}
\usepackage{enumerate}
\newcommand{\bd}{\begin{document}}
\newcommand{\ed}{\end{document}}
\newcommand{\bc}{\begin{center}}
\newcommand{\ec}{\end{center}}
\newcommand{\vs}{\vspace}
\newcommand{\hs}{\hspace}
\newcommand{\beq}{\begin{equation}}
\newcommand{\eeq}{\end{equation}}
\newcommand{\beqs}{\begin{eqn*}}
\newcommand{\eeqs}{\end{eqn*}}
\newcommand{\bq}{\begin{quote}}
\newcommand{\eq}{\end{quote}}
\newcommand{\lb}{\linebreak}
\newcommand{\mb}{\makebox}
\newcommand{\fb}{\framebox}
\newcommand{\mc}{\multicolumn}
\newcommand{\ben}{\begin{enumerate}}
\newcommand{\een}{\end{enumerate}}
\newcommand{\bit}{\begin{itemize}}
\newcommand{\eit}{\end{itemize}}
\newcommand{\ov}{\overline}
\newcommand{\un}{\underline}
\newcommand{\lt}{\left}
\newcommand{\rt}{\right}
\newcommand{\ba}{\begin{array}}
\newcommand{\ea}{\end{array}}
\newcommand{\beqa}{\begin{eqnarray}}
\newcommand{\eeqa}{\end{eqnarray}}
\newcommand{\beqas}{\begin{eqnarray*}}
\newcommand{\eeqas}{\end{eqnarray*}}
\newcommand{\bfg}{\begin{figure}}
\newcommand{\efg}{\end{figure}}
\newcommand{\pad}{\partial}
\newcommand{\nn}{\nonumber}
\newcommand{\la}{\leftarrow}
\newcommand{\ra}{\rightarrow}
\newcommand{\lgla}{\longleftarrow}
\newcommand{\lgra}{\longrightarrow}
\newcommand{\La}{\Leftarrow}
\newcommand{\Ra}{\Rightarrow}
\newcommand{\Lra}{\Leftrightarrow}
\newcommand{\Lgla}{\Longleftarrow}
\newcommand{\Lgra}{\Longrightarrow}
\renewcommand{\a}{\alpha}
\renewcommand{\b}{\beta}
\newcommand{\g}{\gamma}
\newcommand{\G}{\Gamma}
\renewcommand{\d}{\delta}
\newcommand{\D}{\Delta}
\newcommand{\e}{\epsilon}
\newcommand{\eps}{\epsilon}
\newcommand{\s}{\sigma}
\renewcommand{\l}{\lamda}
\newcommand{\m}{\mu}
\newcommand{\n}{\nu}
\renewcommand{\S}{\Sigma}
\newcommand{\p}{\pi}
\newcommand{\om}{\omega}
\newcommand{\Om}{\Omega}
\newcommand{\tri}{\triangle}
\newcommand{\ti}{\times}
\newcommand{\f}{\frac}
\newcommand{\ds}{\displaystyle}
\newcommand{\bm}[1]{\mb{{\boldmath $#1$}}}
\newcommand{\alter}[2]{\lt\{ \ba{ll}#1 \\ #2 \ea \rt.}
\newcommand{\alt}[4]{\lt\{ \ba{ll}#1 & \mb{if \, \,}#2 \\ #3 & \mb{}#4 \ea
    \rt.}
\newcommand{\altn}[4]{\lt\{ \ba{rl}#1 & \mb{if \, \,}#2 \\ #3 & \mb{}#4 \ea
    \rt.}
\newcommand{\altif}[4]{\lt\{ \ba{ll}#1 & \mb{if \, \,}#2 \\ #3 &
\mb{if \, \,}#4 \ea \rt.}
\newcommand{\altnif}[4]{\lt\{ \ba{rl}#1 & \mb{if \, \,}#2 \\ #3 &
\mb{if \, \,}#4 \ea \rt.}
\newcounter{algc}
\newcounter{romc}
\newcounter{Alphc}
\newcommand{\bl}{\begin{list}{{\it Step} ~\arabic{algc}~:} {\usecounter{algc}
                \setlength{\topsep}{0pt} \setlength{\itemsep}{0pt}}}
\newcommand{\el}{\end{list}}
\newcommand{\blr}{\begin{list}{~\roman{romc}~:} {\usecounter{romc}
                \setlength{\topsep}{0pt} \setlength{\itemsep}{0pt}}}
\newcommand{\elr}{\end{list}}
\newcommand{\bla}{\begin{list}{~\Alph{Alphc}~:} {\usecounter{Alphc}
                \setlength{\topsep}{0pt} \setlength{\itemsep}{0pt}}}
\newcommand{\ela}{\end{list}}
\newcommand{\InGaAs}{In$_{0.53}$Ga$_{0.47}$As}

\newtheorem{theorem}{Theorem}
\begin{document}
\title{Revisiting the theory of ferroelectric negative capacitance}
\author{Kausik Majumdar$^{1\dagger}$, Suman Datta$^2$, and Satyavolu Papa Rao$^3$
\thanks{
$^1$Department of Electrical Communication Engineering, Indian Institute of Science, Bangalore 560012, India.
$^2$Department of Electrical Engineering, University of
Notre Dame, Notre Dame, IN 46556-5637 USA.
$^3$SUNY Poly SEMATECH, 257 Fuller Road, Albany, NY 12203, USA,
$^\dagger$Corresponding author, Email: kausikm@ece.iisc.ernet.in}}
\date{}
\maketitle
\begin{abstract}
In this paper we revisit the theory of negative capacitance, in a (i) standalone ferroelectric, (ii) ferroelectric-dielectric, and (iii) ferroelectric-semiconductor series combination, and show that it is important to minimize the total Gibbs free energy of the combined system (and not just the free energy of the ferroelectric) to obtain the correct states. The theory is explained both analytically and using numerical simulation, for ferroelectric materials with first order and second order phase transitions. The exact conditions for different regimes of operation in terms of hysteresis and gain are derived for ferroelectric-dielectric combination. Finally the ferroelectric-semiconductor series combination is analyzed to gain insights into the possibility of realization of steep slope transistors in a hysteresis free manner.
\end{abstract}
\section{Introduction}
Reducing the supply voltage while maintaining the performance is one of the key focus areas of current device research, which would enable reduction of power consumption up to the system level \cite{tnt10}. It is well known that the long high energy tail of the Fermi-Dirac distribution of carrier population at the source junction does not allow the MOSFET current to be changed by any more than a decade for every 60 mV change in the gate voltage at room temperature. This is a fundamental bottleneck of MOSFET operation that limits the supply voltage scaling. Ways to beat this subthreshold slope limit of 60 mV/decade have been intensely investigated in the past \cite{ion11}-\cite{ss08}.

To this end, ferroelectric negative capacitance FET (FerroFET) was proposed \cite{ss08} where the gate insulator of a MOSFET is replaced by a ferroelectric material. With an increase in the gate voltage (over a certain range), the internal \lq\lq{}negative capacitance\rq\rq{} of the ferroelectric forces the voltage drop across itself to decrease, which in turn increases the channel surface potential of the semiconductor by a value which is more than the change of the gate voltage. Such a gain mechanism between the external gate voltage and the internal channel surface potential allows for a larger change in drain current than what is predicted by 60 mV/decade, even though the current-surface potential relationship is still limited by the tail of the Fermi-Dirac distribution \cite{ss08}-\cite{sdUn}.

One question that is frequently asked in this context is whether it is possible to maintain such voltage gain in a hysteresis free way so as to achieve the eventual goal of reducing supply voltage of digital logic. In the recent past, there have been a number of efforts to investigate this, both theoretically \cite{ss08}-\cite{jain14} and experimentally \cite{khan11apl}-\cite{appleby14}. However, a clear understanding of the mechanism of such devices is still lacking in the literature, which is partly due to the solution methodology typically adopted - by minimizing the Gibbs free energy for the ferroelectric and then equating the ferroelectric polarization at the minimum energy point to the charge per unit area of the series capacitance \cite{ss08}-\cite{jain14}. Unfortunately, this does not necessarily minimize the total Gibbs free energy of the combined system since both components (ferroelectric and series capacitance) can have complex dependence of free energy on appropriate state variable. It turns out that although both the approaches gives rise to same results for a linear capacitor placed in series with the ferroelectric, the results can be quite different when the series capacitance is non-linear in nature (for example, semiconductor channel). Our aim in this paper is threefold: (i) to establish a theory based on the minimization of Gibbs free energy of the {\it whole} system, (ii) to find the exact conditions for hysteresis free gain in a ferroelectric-dielectric series combination to check if it is possible to have a design window for such operation, and (iii) to understand a FerroFET operation using a one dimensional FerroMOSCAP analysis to elucidate whether a sub-60 mV/decade operation is possible in a hysteresis free manner.

The rest of the paper is organized as follows: The method of Gibbs free energy minimization is established using a simple example of two linear dielectric capacitors in series in sec. \ref{sec:C1_C2}. The concept of negative capacitance is then explained using a single standalone ferroelectric capacitor in sec. \ref{sec:F}. This is followed by a detailed analysis of ferroelectric-dielectric series combination in sec. \ref{sec:F_C}. The FerroMOSCAP analysis is performed in sec. \ref{sec:F_S}, which is followed by discussion on some practical aspects in sec. \ref{sec:discuss}. Conclusions that can be drawn are presented in sec. \ref{sec:conclude}.
\section{Method of Gibbs free energy minimization}\label{sec:C1_C2}
When a system is excited by an external stimulus $X$, and if $Y$ is an appropriate internal state variable, the system reorganizes $Y$ in such a manner that the Gibbs free energy of the whole system is minimized.
\begin{figure}[!hbt]
\centering
\includegraphics[scale=0.3] {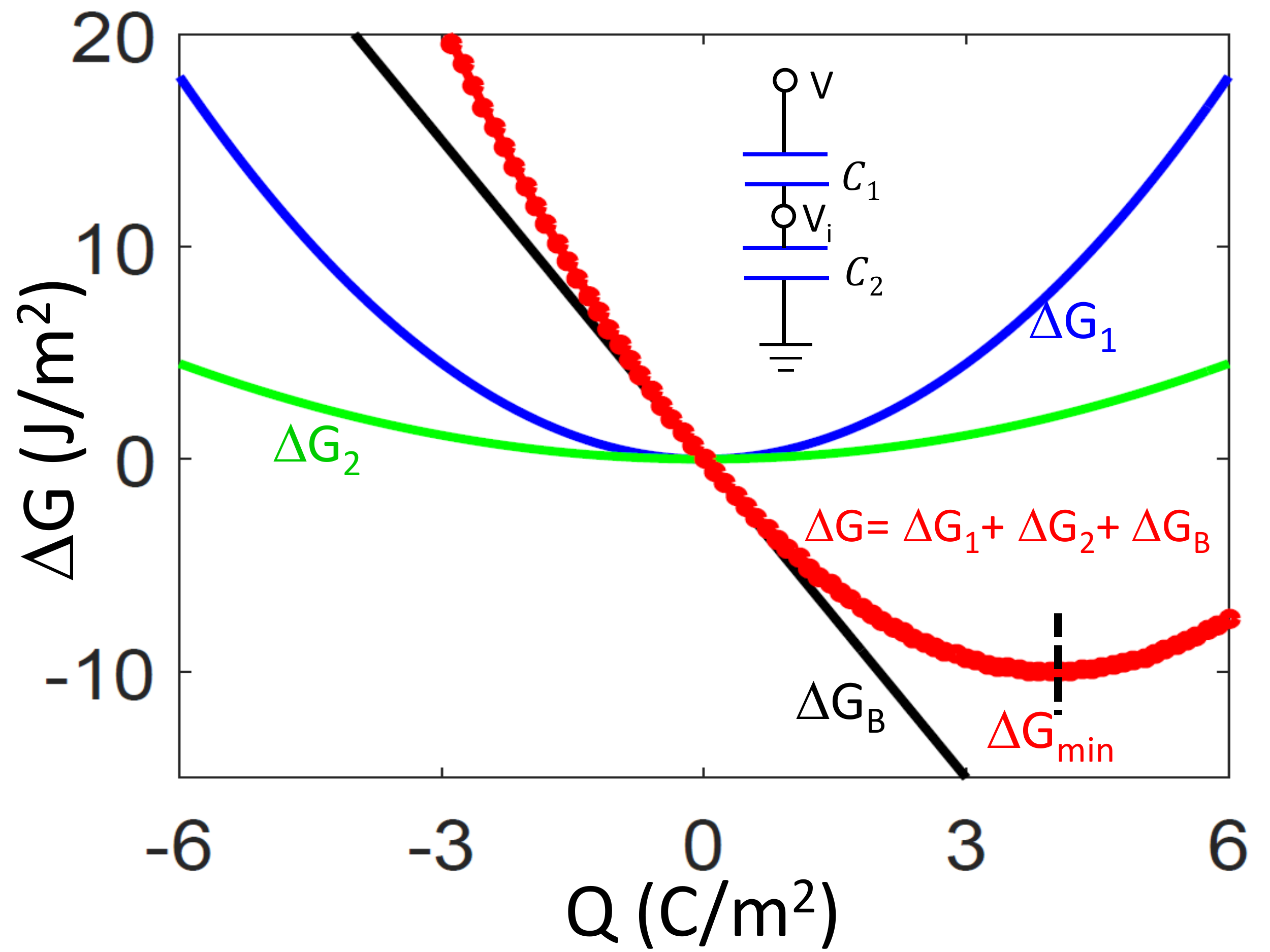}
\vspace{-0.1in}
\caption{Gibbs free energy ($\Delta G$) of two capacitors ($1$ and $4$ F/m$^2$ in series, with an applied bias $V=5$ V. The individual Gibbs free energy of the two capacitors ($\Delta G_{1,2}$) and the battery ($\Delta G_B$) are also shown. The minima of $\Delta G_{1,2}$ do not coincide with total free energy minimum.}\label{fig:C1_C2}
\end{figure}
To explain this, we use a simple example of two capacitors, with $C_1$ and $C_2$ being the capacitance per unit area, connected in series and excited by an external voltage $V$, as shown in the inset of Fig. \ref{fig:C1_C2}. For isothermal process and in the absence of any stress, the Gibbs free energy ($\Delta G$) of a dielectric capacitor is just the electrostatic energy $\int \bar{D}\cdot d\bar{E}$, where $\bar{D}$ is the displacement vector and $\bar{E}$ is the electric field. In this paper, we assume all quantities to be uniform in-plane and perform one dimensional analysis, removing the vector signs for simplicity. $D$ then becomes equal to the charge per unit area ($Q$) at the capacitor plates. The $\Delta G$ of the combined system is given by
\beq
\Delta G = \frac{Q^2}{2C_1} + \frac{Q^2}{2C_2} - QV
\eeq
where $V$ is the external supply voltage.
For a given $V$, the system will settle $Q$ in such a way that $\Delta G$ is minimized, i.e. $\frac{\partial \Delta G}{\partial Q}=0$ and $\frac{\partial^2\Delta G}{\partial Q^2}>0$, which gives rise to the well known result of combined capacitance $C=\frac{Q}{V}=\frac{C_1}{C_1+C_2}$. The same result is readily obtained electrostatically by equating the charges on the capacitor plates, i.e. $C_1(V-V_i)=C_2V_i$ where $V_i$ is the internal node voltage. It\rq{}s interesting to note from  Fig. \ref{fig:C1_C2} that when $\Delta G$ is minimized, both $\Delta G_1$ and $\Delta G_2$ are above their individual minimum due to non-monotonic dependence of $\Delta G_{1,2}$ on $Q$. This leads to an important conclusion that when a system stabilizes in its Gibbs free energy minimum, the subsystems may not necessarily be in their individual energy minimum with respect to the internal state variable.
\section{A standalone ferroelectric capacitor}\label{sec:F}
\begin{figure}[!hbt]
\centering
\includegraphics[scale=0.35] {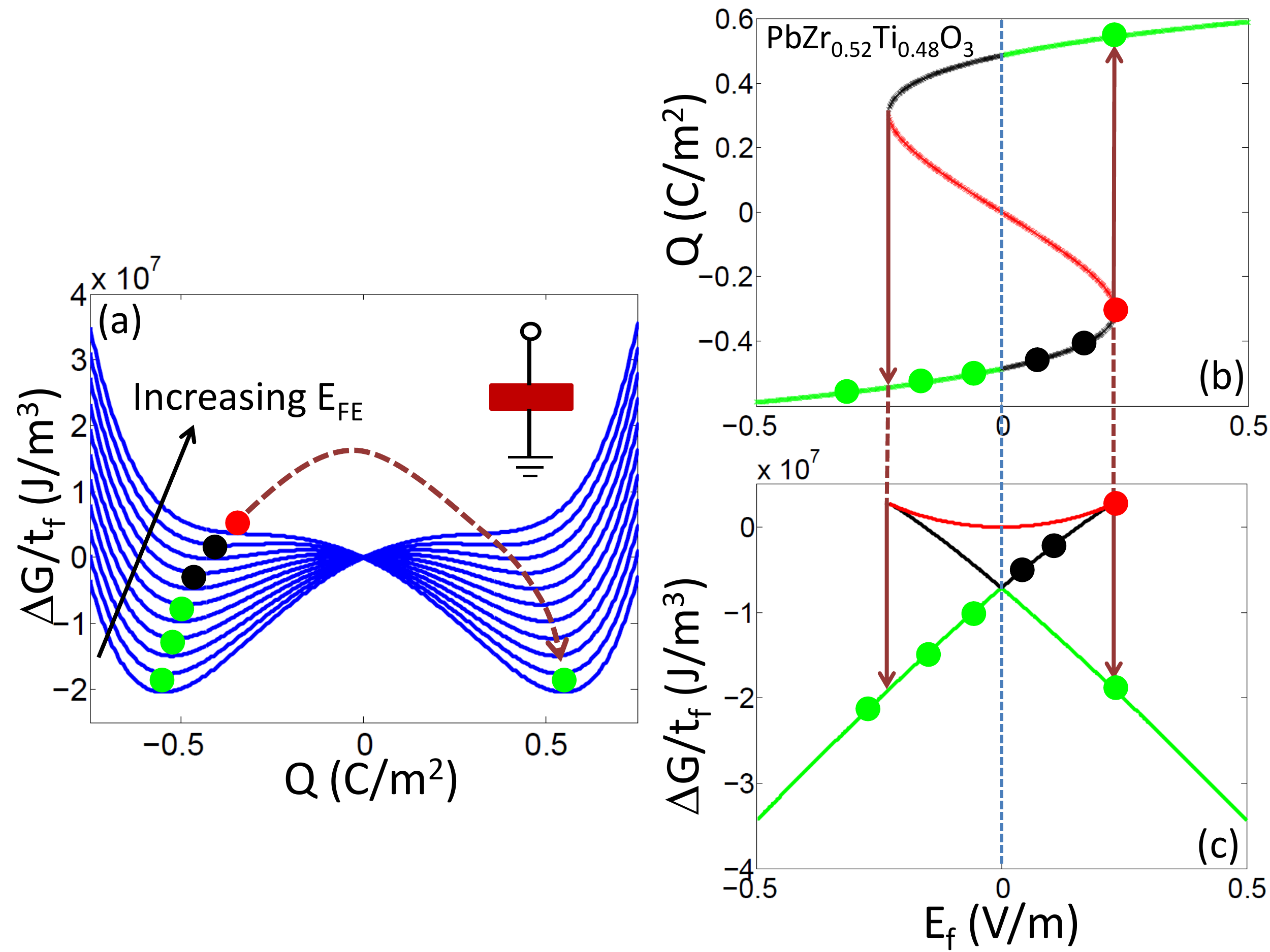}
\vspace{-0.2in}
\caption{(a): $\Delta G$ of a standalone Pb$_{0.52}$Zr$_{0.48}$TiO3 (PZT) capacitor, plotted as a function of $Q$ for varying electric field. (b): $Q$ and (c): $\Delta G$, plotted as a function of electric field across it. The green, black and red dots show absolutely stable, metastable and unstable states.}\label{fig:F_alone}
\end{figure}

Using the phenomenological treatment of Landau, and ignoring any surface and domain boundary effect, the Gibbs free energy of the ferroelectric capacitor is given by \cite{lgbook}
\beq\label{eq:del_Gf}
\Delta G_f = t_f\times(\alpha_1Q^2 + \alpha_{11}Q^4 + \alpha_{111}Q^6 - E_fQ)
\eeq
where $E_f$ is the constant field within the ferroelectric and is an independent excitation variable. $Q$ is charge per unit area of the capacitor plate, and $\alpha_i$ are the Landau coefficients of the ferroelectric and are summarized in Table \ref{tab:ferro_params}  for PbZr$_{0.48}$Ti$_{0.52}$O$_3$ (PZT) \cite{rabebook} and BaTiO$_3$ (BTO) \cite{wangJap07}. Using $\frac{\partial\Delta G_f}{\partial Q}=0$, we find
\beq\label{eq:Ef}
E_f = 2\alpha_1Q + 4\alpha_{11}Q^3 + 6\alpha_{111}Q^5
\eeq
\begin{table}[!hbt]
\caption{List of Landau parameters for PZT and BTO at $T$$=$300K.}\label{tab:ferro_params}
\vs{-0.2in}
\bc
\begin{tabular}{|c|c|c|}
\hline
\textbf{Parameters} & \textbf{PZT} \cite{rabebook} & \textbf{BTO} \cite{wangJap07}\\
\hline
$\alpha_{1}$ (mF$^{-1}$) & $-5.2393\times 10^7$ & $-3.2851\times 10^7$\\
\hline
$\alpha_{11}$ (m$^5$F$^{-1}$C$^{-2}$) & $5.8252\times 10^7$ & $-6.300\times10^8$\\
\hline
$\alpha_{111}$ (m$^9$F$^{-1}$C$^{-4}$) & $1.5039\times 10^8$ & $4.3000\times 10^9$\\
\hline
\end{tabular}
\ec
\end{table}
The results are summarized for PZT in Fig. \ref{fig:F_alone}, which shows that as $E_f$ is increased from a large negative value, the free energy of the ferroelectric increases, and the system gradually moves from stable (green dot) to metastable state (black dot). On further increase of electric field, the ferroelectric remains in the metastable state as long as there is an energy barrier between the current state and the newly created stable state. Finally, the ferroelectric reaches the red dot point at an electric field equal to the coercive field, where all energy barrier is removed and the polarization of the ferroelectric abruptly changes as it moves to the other free energy minimum. In the process of such a hysteretic polarization jump, the ferroelectric exhibits a transient negative capacitance as it moves through the states where $\frac{\partial^2 \Delta G}{\partial Q^2}<0$. However, a standalone ferroelectric does not traverse the negative slope portion in the $Q$-$E_f$ curve [red solid line in Fig. \ref{fig:F_alone}(a)-(b)], since this line corresponds to the locus of the local maxima in the free energy landscape where the ferroelectric is not allowed to stabilize.
\section{A ferroelectric capacitor in series with a constant capacitor}\label{sec:F_C}
\begin{figure}[!hbt]
\centering
\includegraphics[scale=0.4] {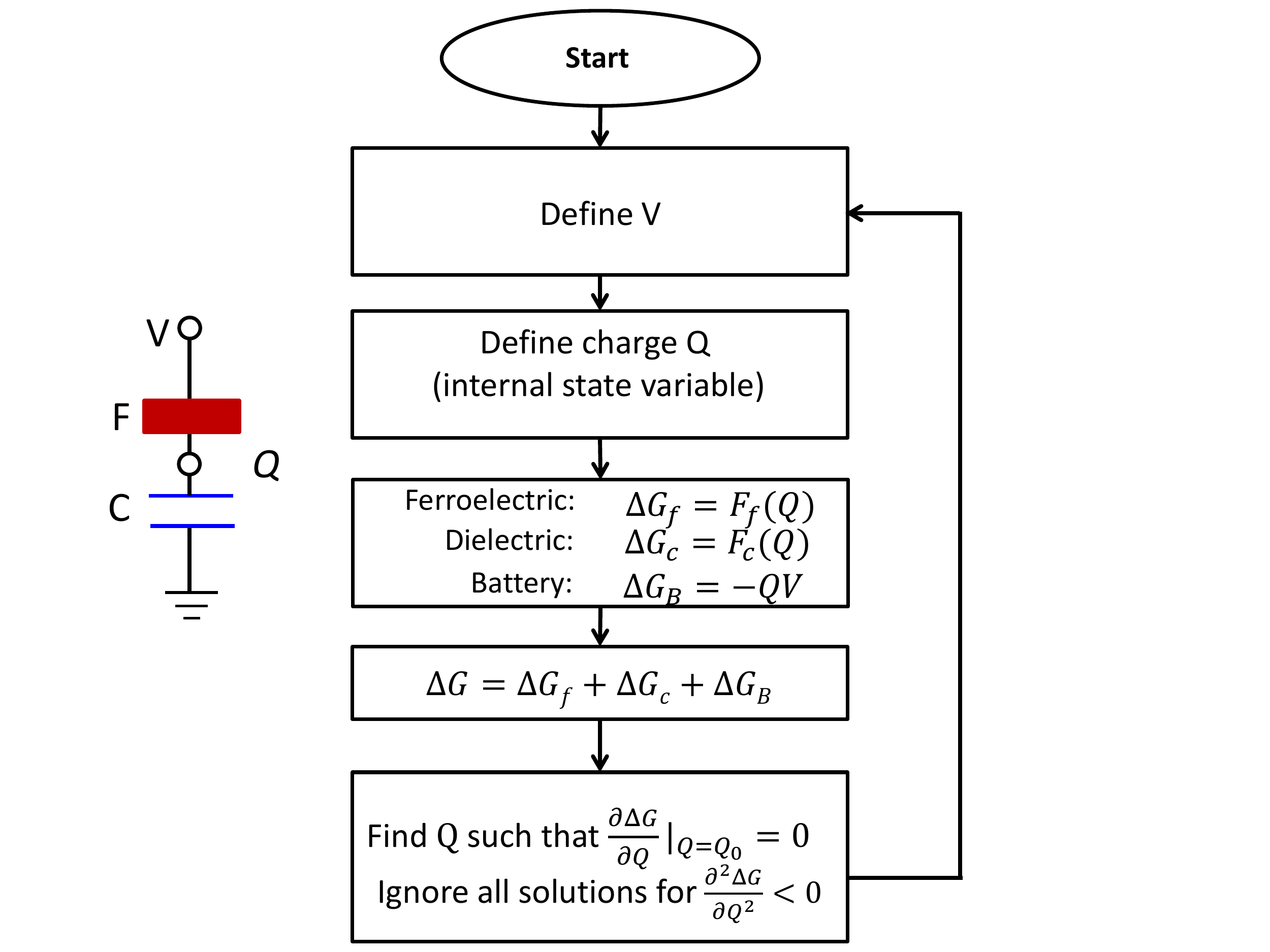}
\caption{Flow chart explaining the numerical simulation methodology to find states for a combined ferroelectric-dielectric system.}\label{fig:flow_chart}
\end{figure}

This situation is schematically shown in Fig. \ref{fig:flow_chart}.
It has been proposed in the past \cite{ss08} that the negative capacitance can be accessed in a hysteresis free way by adding a dielectric capacitor in series. Such a combination has been generally analyzed \cite{ss08}-\cite{jain14} by first minimizing the Gibbs free energy of the ferroelectric assuming the electric field within the ferroelectric ($E_f$) is the independent excitation, and then equating the polarization at the ferroelectric free energy minimum to the charge per unit area of the capacitor plate. Such an approach does not necessarily minimize the total $\Delta G$ of the ferroelectric/dielectric combination.

Before discussing further we state few assumptions that we take in the analysis presented in the rest of the paper: (i) The ferroelectric is perfectly uniform and is governed by the bulk expression in Eq. \ref{eq:del_Gf}. Any relaxation due to domain boundaries is ignored. (ii) The ferroelectric-dielectric has a perfect interface with no added strain or charge. (iii) The leakage current through the whole stack is negligible.

The Gibbs free energy of the ferroelectric/dielectric combination under applied bias $V$ is given by
\beq\label{eq:del_G}
\Delta G = t_f\Big{[}\alpha_1Q^2 + \alpha_{11}Q^4 + \alpha_{111}Q^6 \Big{]} + \frac{Q^2}{2C} - QV
\eeq
where $t_f$ is the thickness of the ferroelectric film, $C=\eps_0\eps_c/t_c$, $\epsilon_0$ is the permittivity of free space, $\epsilon_c$ is dielectric constant, and $t_c$ is the thickness of the dielectric film.
Using $\frac{\partial\Delta G}{\partial Q} = 0$, we get
\beq\label{eq:d1}
V=(2\alpha_1t_f + \frac{1}{C})Q + 4\alpha_{11}t_fQ^3 + 6\alpha_{111}t_fQ^5
\eeq
Note that the condition in Eq. \ref{eq:d1} only dictates that $\Delta G$ is at an extremum. To make sure that $\Delta G$ is minimized, we need $\frac{\partial^2\Delta G}{\partial Q^2} > 0$, which in turn  confirms that the net differential capacitance ($C_T$) of the whole system is positive since $C_T=[\frac{\partial^2\Delta G}{\partial Q^2}]^{-1}$. We can solve Eq. \ref{eq:d1} to obtain the states of the combined system at a given $V$.
\subsection{Condition for hysteresis-free operation}
Noting that the primary cause for hysteresis in a system is the existence of two free energy minima separated by an energy barrier, we argue that the sufficient condition for hysteresis-free operation is that the system should have only one free energy minimum. Mathematically, this would mean that the $\Delta G$-$Q$ curve (which is a polynomial of degree $6$) will have less than two points of inflection, or the equation
\beq\label{eq:d2}
\frac{\partial^2\Delta G}{\partial Q^2}=0
\eeq
will have less than two solutions for real $Q$. Using Eq. \ref{eq:del_G}, Eq. \ref{eq:d2} becomes
\beq
Q^4 + \frac{2\alpha_{11}}{5\alpha_{111}}Q^2 + \frac{1}{30\alpha_{111}}(2\alpha_1+\frac{1}{Ct_f}) = 0
\eeq
which gives
\beq\label{eq:D2}
Q^2 = \frac{1}{2}\Bigg{[}-\frac{2\alpha_{11}}{5\alpha_{111}} \pm \sqrt{\Big{(}\frac{2\alpha_{11}}{5\alpha_{111}}\Big{)}^2 - \frac{2}{15\alpha_{111}}\Big{(}2\alpha_1+\frac{1}{Ct_f}\Big{)}}\Bigg{]}
\eeq
\subsubsection{Case I - Second order phase transition ($\alpha_{11}>0$, $\alpha_{111}>0$)}
From Eq. \ref{eq:D2}, by observation, $Q$ has no real solution for
\beq
\frac{1}{Ct_f} > -2\alpha_1
\eeq
which corresponds to absence of points of inflection, leading to hysteresis-free operation with a single minimum. As a sanity check, this condition also ensures that the coefficient of $Q$ in Eq. \ref{eq:d1} is positive. On the other hand, for $\frac{1}{Ct_f} < -2\alpha_1$, $Q$ will have only $2$ real solutions corresponding to conventional bistable $\Delta G$-$Q$ curve.
\subsubsection{Case II - First order phase transition ($\alpha_{11}<0$, $\alpha_{111}>0$)}
(i) For no real solution of $Q$, the term under the square root is less than zero, which reduces to
\beq
\frac{1}{Ct_f} > -2\alpha_1 + \frac{6\alpha_{11}^2}{5\alpha_{111}}
\eeq
for hysteresis free operation.\\
(ii) For only $2$ real solutions of $Q$ (which correspond to bistable states as in case I),
\beq\nonumber
-\frac{2\alpha_{11}}{5\alpha_{111}} - \sqrt{\Big{(}\frac{2\alpha_{11}}{5\alpha_{111}}\Big{)}^2 - \frac{2}{15\alpha_{111}}\Big{(}2\alpha_1+\frac{1}{Ct_f}\Big{)}} < 0
\eeq
which reduces to
\beq
\frac{1}{Ct_f} < -2\alpha_1
\eeq
which is exactly the same condition we obtained for $\alpha_{11}>0$.\\
(iii) For $-2\alpha_1 < \frac{1}{Ct_f} < -2\alpha_1 + \frac{6\alpha_{11}^2}{5\alpha_{111}}$, there exist $4$ points of inflection for real values of $D$, which correspond to three minima in $\Delta G$ leading to two hysteresis windows.
\subsection{Condition for differential gain}
For enhancement in capacitance (or differential voltage gain), we must have $C_T >C$. Again using the fact that $C_T=[\frac{\partial^2\Delta G}{\partial D^2}]^{-1}$,  we obtain the following inequality:
\beq\label{eq:g}
Q^4 + \frac{2\alpha_{11}}{5\alpha_{111}}Q^2 + \frac{\alpha_1}{15\alpha_{111}} < 0
\eeq
This leads to a certain range of operation where we expect to have capacitance gain:
\beq
0 < Q^2 < -\frac{\alpha_{11}}{5\alpha_{111}}+  \sqrt{\Big{(}\frac{\alpha_{11}}{5\alpha_{111}}\Big{)}^2 - \frac{\alpha_1}{15\alpha_{111}}}
\eeq
This same condition holds good for both first and second order ferroelectric materials.

\begin{table*}[!hbt]
\caption{Summary of conditions for different regimes of operation in ferroelectric-dielectric series combination.}\label{tab:conditions}
\vs{-0.1in}
\bc
\begin{tabular}{|c|c|c|}
\hline
\textbf{Regime of operation} & $\mathbf{\alpha_{11}>0, \alpha_{111}>0}$ & $\mathbf{\alpha_{11}<0, \alpha_{111}>0}$\\
\hline
Hysteresis free (single minimum) \textbf{(R1)} & $\frac{1}{Ct_f} > -2\alpha_1$ & $\frac{1}{Ct_f} > -2\alpha_1 + \frac{6\alpha_{11}^2}{5\alpha_{111}}$\\
\hline
Hysteretic (double minima) \textbf{(R2)} & $\frac{1}{Ct_f} < -2\alpha_1$ & $\frac{1}{Ct_f} < -2\alpha_1$\\
\hline
Hysteretic (three minima) \textbf{(R3)} & - &$-2\alpha_1  < \frac{1}{Ct_f} <  -2\alpha_1 + \frac{6\alpha_{11}^2}{5\alpha_{111}}$\\
\hline
\end{tabular}
\ec
\end{table*}

Although we were able to treat the problem analytically owing to simple linear nature of the charge-voltage relation of the dielectric capacitor, for a more general analysis, we need numerical simulation. The steps are shown in Fig. \ref{fig:flow_chart}. At a given external voltage $V$, we choose charge $Q$ as the internal state variable and create a look-up table based on the general charge-Gibbs free energy relation of the generic capacitor $C$ and the ferroelectric. The total $\Delta G$ is then found as  $\Delta G=\Delta G_f+\Delta G_c + \Delta G_B$, where $\Delta G_B$ is the free energy of the power supply. We choose all $Q=Q_0$ for which the total Gibbs free energy $\Delta G$ has extrema, of which the minima correspond to either stable or metastable states, and the maxima correspond to the unstable states (hence ignored).

\begin{figure*}[!hbt]
\centering
\includegraphics[scale=0.45] {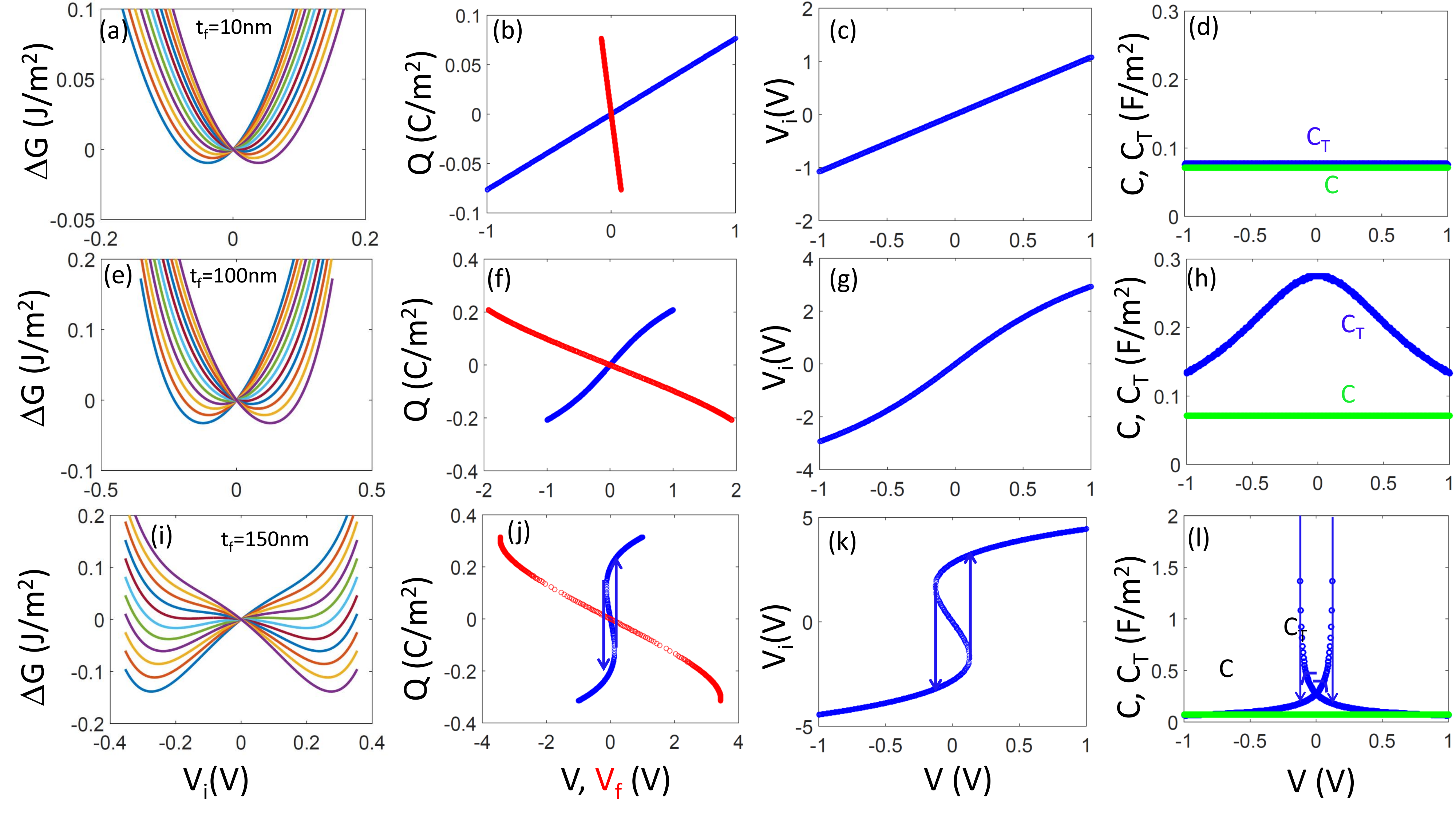}
\caption{Characteristics of varying thickness ferroelectric (PZT) with second order phase transition ($\alpha_{11}>0, \alpha_{111}>0$) in series with $25$ nm thick STO. Top row (a)-(d): $t_f$=$25$nm, middle row (e)-(h): $t_f$=$100$nm, and bottom row (i)-(l): $t_f$=$150$nm. (a), (e) and (i): $\Delta G$ of the whole system, plotted as a function of internal node voltage $V_i$, for different applied bias $V$. (b), (f), (j): $Q$-$V$ and $Q$-$V_f$ are shown in blue and red, respectively, where $V_f$ is the voltage drop across the ferroelectric. In (b) and (f), although the PZT exhibits negative capacitance (in red), the overall capacitance remains positive (in blue) due to the single minimum character of the $\Delta G$. There is an unstable portion of the $Q$-$V$ curve in (j) resulting in hysteretic jumps. (c), (g), (k): The internal node voltage $V_i$ is shown as a function of input voltage $V$. (d), (h), (l): The capacitance ($C$) of the STO and the total capacitance ($C_T$) of the PZT/STO stack.}\label{fig:F_C_PZT}
\end{figure*}

Figures \ref{fig:F_C_PZT} summarizes the results for varying thickness of PZT (with $\alpha_{11}>0$) deposited on $25$nm thick SrTiO$_3$ (STO). Dielectric constant ($\epsilon_c$) of STO is assumed to be $200$. The thickness of the PZT layer in the top, middle and bottom rows are $10$nm, $100$nm and $150$nm, respectively. We clearly see that the total $\Delta G$ maintains its single minimum character for top and middle rows, leading to hysteresis free operation. The corresponding charge per unit area ($Q$) plot in Fig. \ref{fig:F_C_PZT}(b) and (f) indicate that the PZT goes through negative capacitance (negative slope of the $Q$-$V_f$ plot in red where $V_f$ is the voltage drop across the ferroelectric) regime, although the slope for the whole system ($Q$-$V$ plot) remains positive (in blue), in agreement with the single minimum $\Delta G$ plot in Fig. \ref{fig:F_C_PZT}(a) and (e). When the ferroelectric layer is very thick, $\Delta G_c$ cannot completely compensate the double minimum behavior of $\Delta G_f$ [Fig. \ref{fig:F_C_PZT}(i)] and thus the combined system shows hysteretic character as observed in Fig. \ref{fig:F_C_PZT} (j). In Fig.  \ref{fig:F_C_PZT}(c), (g) and (k), the corresponding internal node voltages $V_i$ are plotted as a function of external voltage $V$ showing quasi-linear dielectric, hysteresis-free  gain, and hysteretic behavior, respectively. This point is further clarified in the differential capacitance plots in the last column (d, h, and l), where the total capacitance ($C_T$) becomes larger than the dielectric series capacitance ($C$) in a hysteresis-free way in (d) and (h), and with hysteresis in (l).

For the case of $\alpha_{11}<0$, we choose the example of varying thickness of BaTiO$_3$ (BTO) on $25$nm STO, and the results are summarized in Fig. \ref{fig:F_C_BTO} based on Landau parameters listed in Table \ref{tab:ferro_params} from ref. \cite{wangJap07}. We  clearly observe three different regimes of operation, in accordance with Table \ref{tab:conditions}. One important difference from the $\alpha_{11}>0$ case is that the gain region splits away symmetrically from $V=0$ and appears at some nonzero $\pm V$. With larger $t_f$, correspondingly two hysteresis windows open up, and eventually with further increment of $t_f$, they converge to a single large hysteresis window.
\subsection{Peak capacitance in hysteresis-free operation}
\begin{figure}[!hbt]
\centering
\includegraphics[scale=0.35] {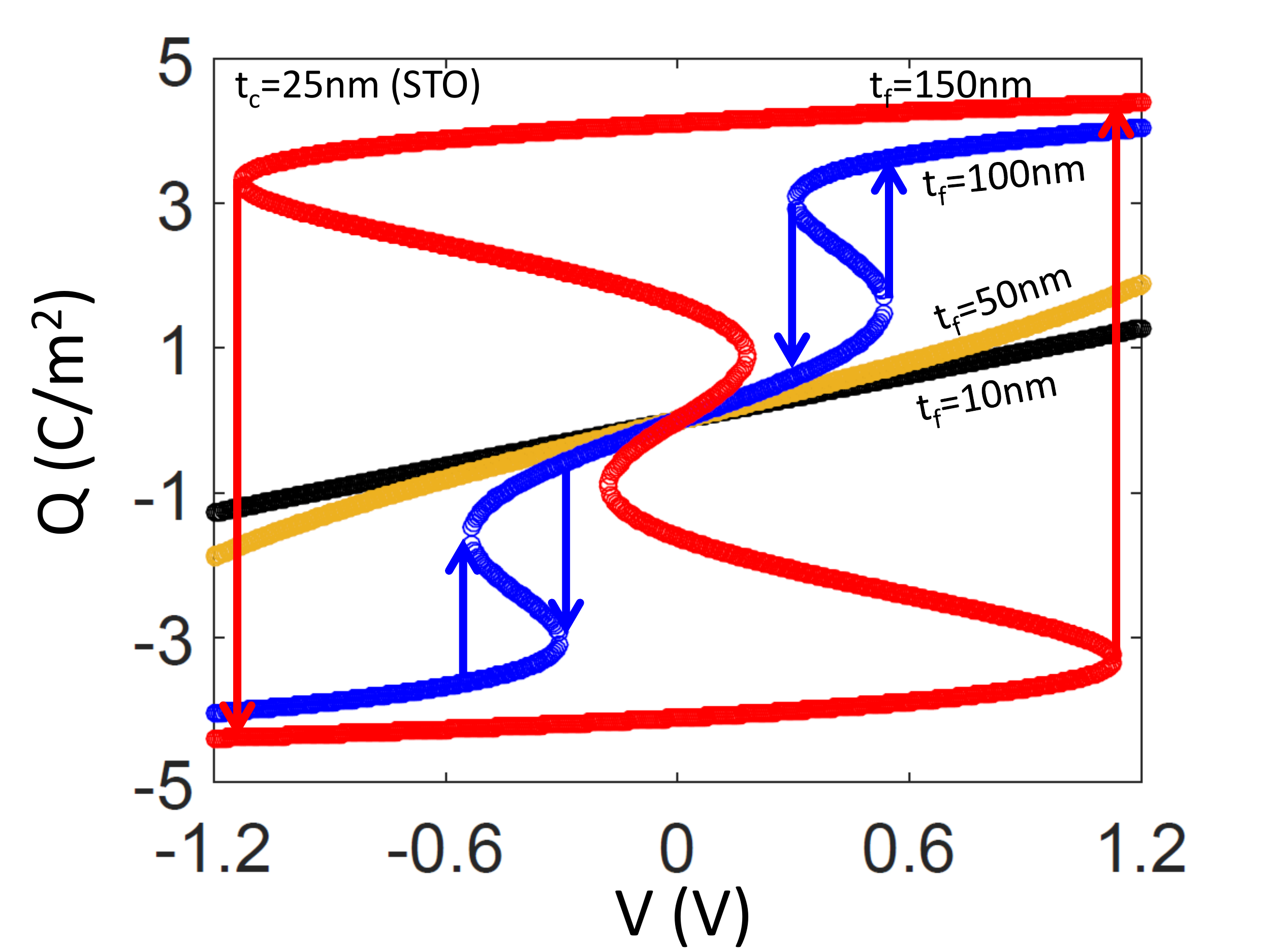}
\caption{$Q$-$V$ characteristics of a varying thickness ferroelectric (BTO) with first order phase transition ($\alpha_{11}<0, \alpha_{111}>0$), placed in series with $25$nm thick STO showing transition from hysteresis-free to double-hysteresis window to single hysteresis window regimes.}\label{fig:F_C_BTO}
\end{figure}

In the case of hysteretic jump, there is an abrupt change in charge, which gives rise to infinite differential capacitance at the steep jump point. On the other hand, in the case of hysteresis free operation, we can find the peak capacitance by noting that the total capacitance $C_T$ as
\beq
\frac{1}{C_T} = \frac{\partial^2\Delta G}{\partial Q^2} = t_f(30\alpha_{111}Q^4 +12\alpha_{11}Q^2 + 2\alpha_1)+\frac{1}{C}
\eeq
Clearly, $C_T$ is maximum when
\beq
Q(5\alpha_{111}Q^2 + \alpha_{11})=0
\eeq
which is satisfied for either $Q=0$ or $Q=\pm(-\frac{\alpha_{11}}{5\alpha_{111}})^{1/2}$. Noting that for $C_T$ to be maximum, we need $\frac{\partial^2}{\partial Q^2}(\frac{1}{C_T})=360\alpha_{111}t_f + 24\alpha_{11}t_f>0$, we easily find that for second order ferroelectric ($\alpha_{11}>0$), we obtain maximum $C_T$ at $Q=0$. On the other hand, for first order ferroelectric ($\alpha_{11}<0$), the maximum is reached for $Q=\pm(-\frac{\alpha_{11}}{5\alpha_{111}})^{1/2}$.
Clearly, $C_T$ is single peak function for second order ferroelectric materials where, while for first order ferroelectric materials, it exhibits a double peak characteristics at symmetric $V$. This is in agreement with our observation in Fig. \ref{fig:F_C_PZT}(h) and also in Fig. \ref{fig:F_C_BTO}.
The corresponding maximum capacitance is given by
\begin{equation*}
C_T^{max} = \left\{
\begin{array}{ll}
\Big{[}\frac{1}{C} + 2\alpha_1t_f \Big{]}^{-1} & \text{if }\alpha_{11} > 0,\\
\Big{[}\frac{1}{C} + 2\alpha_1t_f - \frac{6\alpha_{11}^2}{5\alpha_{111}}t_f \Big{]}^{-1} & \text{if } \alpha_{11} < 0
\end{array} \right.
\end{equation*}
which is in excellent agreement with the simulation predicted peak capacitance in Fig. \ref{fig:F_C_PZT}(h).
\section{A ferroelectric capacitor in series with a semiconductor (FerroMOSCAP)}\label{sec:F_S}
\begin{figure*}[!hbt]
\centering
\includegraphics[scale=0.45] {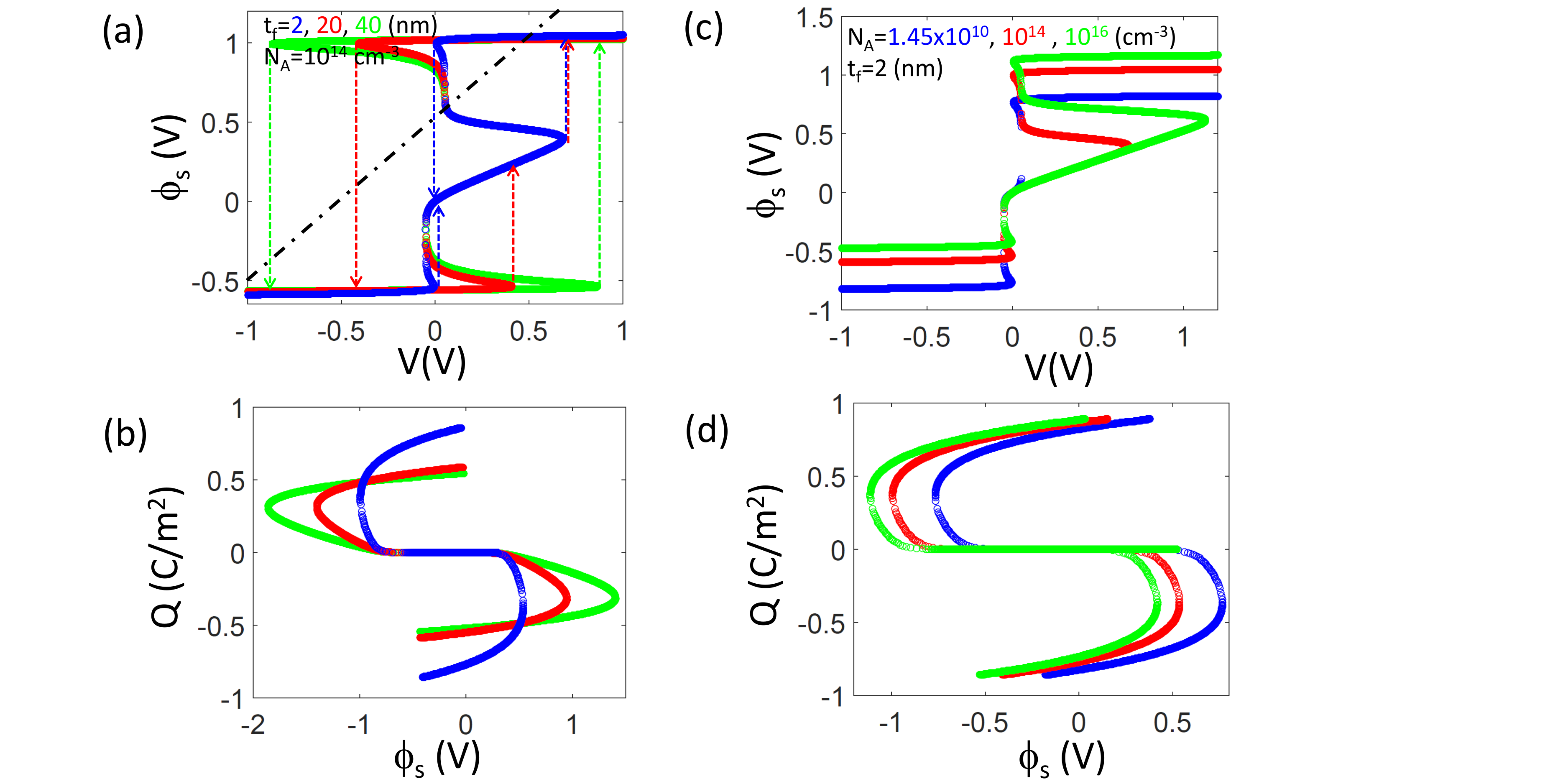}
\caption{(a): Surface potential ($\phi_s$) versus applied gate bias ($V$) and (b): charge per unit area ($Q$) versus surface potential ($\phi_s$) for PZT/Si stack with different PZT thickness $t_f$. $N_A$ is kept fixed at $10^{14}$ cm$^{-3}$. Any gate leakage due to low band offset between PZT and Si has been neglected. The negative slopes in all the plots correspond to local maxima in the total $\Delta G$ of the system (shown in Fig. \ref{fig:F_MOS_delG}) and hence unstable, forcing hysteretic jumps shown by the dotted arrows. The black dashed-dotted lines in (a) indicate unity gain (corresponding to 60 mV/decade). (c)-(d): $t_f$ is kept fixed at $2$nm, and doping density in Si is varied as intrinsic, $10^{14}$ cm$^{-3}$ and $10^{16}$ cm$^{-3}$.}\label{fig:F_MOS_P}
\end{figure*}

We now turn our attention to what happens when a semiconductor is in series with a ferroelectric in a MOSCAP configuration. The primary difference with the previous case is that the areal charge density in the semiconductor is no more a linear function of voltage, rather a more complex function given by \cite{taurbook}:
\beq
Q=\pm \sqrt{2\epsilon_sk_BTN_A}\Big{[}\Big{(}e^{-\phi_s^\prime} + \phi_s^\prime -1\Big{)}
 + \frac{n_i^2}{N_A^2}\Big{(}e^{\phi_s^\prime} - \phi_s^\prime -1\Big{)}\Big{]}^{1/2}
\eeq
where $\phi_s^\prime = \frac{q\phi_s}{k_BT}$, $\phi_s$ is the surface potential, $\epsilon_s$ is the dielectric constant of the semiconductor, $k_B$ is the Boltzmann constant, $T$ is the temperature, $N_A$ is the doping density and $n_i$ is the intrinsic carrier concentration. We use similar method as described in the flow chart in Fig. \ref{fig:flow_chart} for numerical simulation of the FerroMOSCAP structure.
The fact that this charge is significantly less when compared with the typical charge across the ferroelectric for a wide range of input voltage, the charge-voltage relationship becomes distorted for the whole system. The semiconductor surface potential ($\phi_s$) and the charge per unit area ($Q$) of the ferroMOSCAP are shown in Fig. \ref{fig:F_MOS_P} as a function of gate voltage $V$ for different semiconductor doping ($N_A$) and PZT thickness ($t_f$) combinations. The system may encounter different unstable regions, which correspond to the local maxima in total $\Delta G$ and we explicitly show them in Fig. \ref{fig:F_MOS_P} by the negative slope regions in $\phi_s$. Such instabilities are accompanied by a hysteretic jump in $\phi_s$, as indicated by the arrows. Larger substrate doping slows down the building up of charge in the system (with $V>0$) which in turn results in wider hysteresis window. On the other hand, thickness of the ferroelectric controls the return path, with larger hysteresis window for larger $t_f$, as expected. It\rq{}s important to note that, for simplicity, we have assumed same Landau parameters for different thickness of PZT in Fig. \ref{fig:F_MOS_P}.

\begin{figure*}[!hbt]
\centering
\includegraphics[scale=0.4] {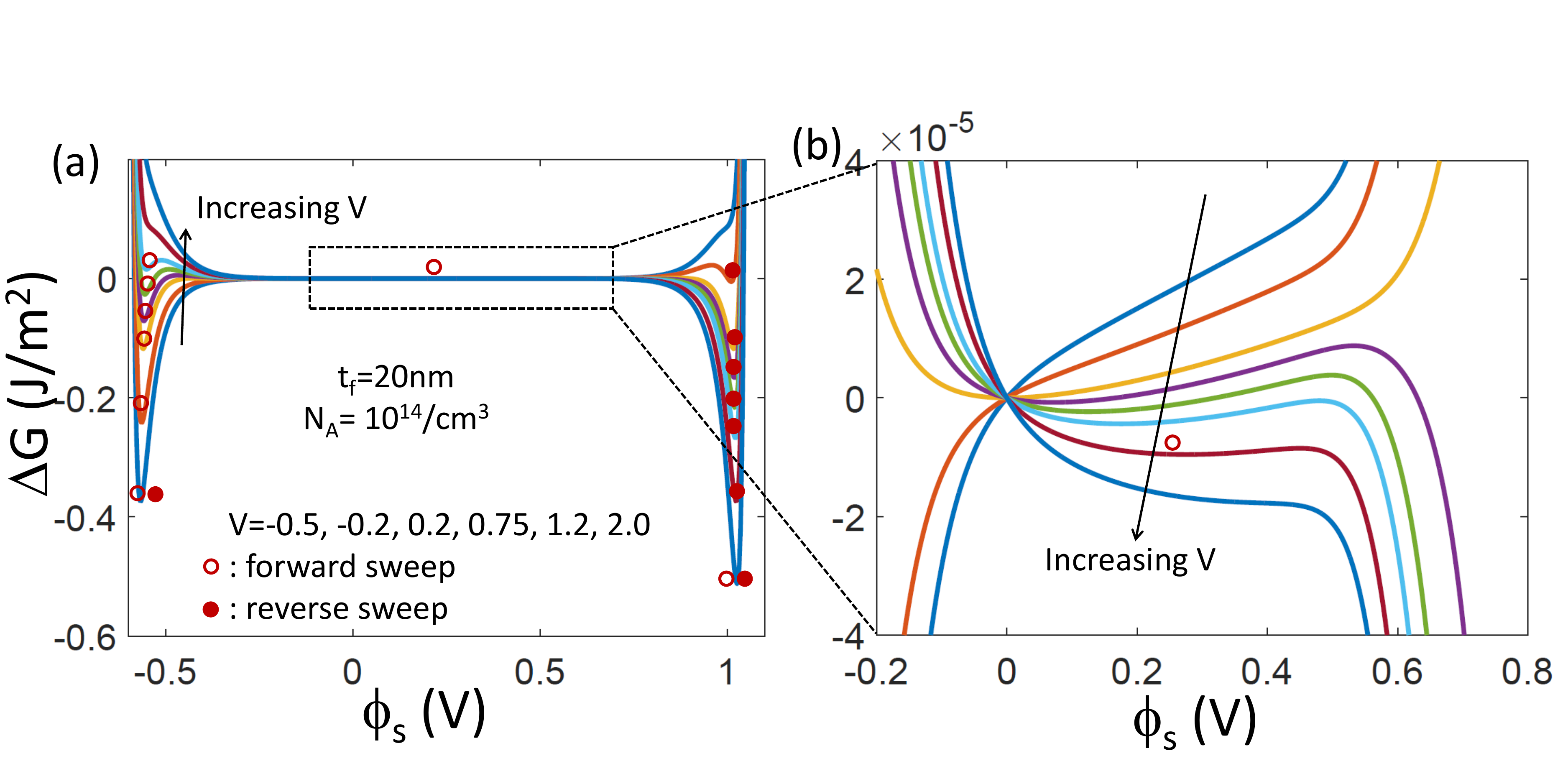}
\vspace{-0.1in}
\caption{Total $\Delta G$ of the PZT/Si stack [with same conditions as the red curves in Fig. \ref{fig:F_MOS_P}(a), (b)] plotted as a function of surface potential $\phi_s$ for different gate voltage $V$. The relatively flat portion in (a) is zoomed in (b) to show the local minima states which control the system as low gate bias. The open and solid circles represent the state of the system in forward and reverse sweep, respectively, which is in agreement with Fig. \ref{fig:F_MOS_P}(a) and (b).}\label{fig:F_MOS_delG}
\end{figure*}

To gain insights into such $Q$-$V$ characteristics, in Fig. \ref{fig:F_MOS_delG}, we show the $\Delta G$ of the whole system as a function of $\phi_s$ for similar conditions as the red traces in Fig. \ref{fig:F_MOS_P}(a),(b). The two strong minima on the left and the right in Fig. \ref{fig:F_MOS_delG}(a) are primarily governed by the strong ferroelectric polarization. However, the seemingly flat portion in the energy landscape can actually possesses a number of local minima [Fig. \ref{fig:F_MOS_delG}(b)], which are governed by the way the charge is modulated in the semiconductor. In particular, as $V$ is increased from a large negative value, the system stabilizes itself at the accessible minimum in $\Delta G$ for every $V$, and $\phi_s$ is traced accordingly. This is explained in Fig. \ref{fig:F_MOS_delG} by the open (forward sweep) and closed (reverse sweep) circles, which correspond to the sweeps indicated in red in Fig. \ref{fig:F_MOS_P}(a) and (b).
\section{Discussion}\label{sec:discuss}
Before we conclude, let us discuss some important points:

(i) \textbf{Effect of domains}: In the whole analysis presented in this paper, we ignored any domain and interface effect, which in reality can significantly perturb the device characteristics. For example, domain nucleations may result in non-uniform feedback from the dielectric in series and hence the conditions for the different regimes of operation as tabulated in Table \ref{tab:conditions} may not be met exactly. As shown in \cite{d2000}, thinning down the ferroelectric may be helpful to reduce such non-uniformity.

(ii) \textbf{Leakage issues}: When a dielectric capacitor is placed in series with a ferroelectric, the ferroelectric forces a large amount of charge across the dielectric, which can easily drive the dielectric to operate close to breakdown. For example, with a ferroelectric polarization of $50\mu$C/cm$^2$ and dielectric $\epsilon_c$ of $50$, the field across the dielectric capacitor is $\sim$$10$MV/cm. A lower $\epsilon_c$ or higher polarization will make the field even higher. This may increase the  leakage, particularly when the ferroelectric has a relatively low barrier offset.


(iii) \textbf{Region of operation}: For a FerroFET, in the subthreshold regime, the semiconductor offers a very small capacitance and provides a large negative feedback to the ferroelectric and it becomes difficult to extract an overall gain between the surface potential and the external gate voltage. Thus sub-60 mV/decade operation at subthreshold may be difficult to achieve in such a configuration. However, when the transistor is in inversion, the effective capacitance of the semiconductor increases significantly providing a less negative feedback to the ferroelectric. Thus above threshold voltage, the charge of a FerroFET may be increased faster than typically achieved in a conventional MOSFET.  This can effectively reduce the required overdrive voltage in a transistor helping to reduce the overall supply voltage.

(iv) \textbf{Requirement of low polarization ferroelectric}: From Fig. \ref{fig:F_MOS_P}, it is clear that a complete hysteresis-free operation with sufficient gain may be difficult to achieve under realistic doping conditions and ferroelectric thickness. To reduce the width of the hysteresis window, it is important to look for very low polarization ferroelectric which will allow the semiconductor with its relatively low charge to control the Gibbs free energy landscape of the ferroelectric-semiconductor combination. The in-built depolarization field in the ferroelectric film at reduced thickness can also help to achieve this.
\section{Conclusion}\label{sec:conclude}
In conclusion, we have investigated three different configurations, namely (i) a standalone ferroelectric, (ii) a ferroelectric in series with a dielectric capacitor, and (iii) a ferroelectric in series with a semiconductor. We pointed out that in all the cases, it is important to minimize the total Gibbs free energy of the whole system (and not just the ferroelectric) to reach to the correct solution for the states. By using this methodology, we found that it is possible to achieve hysteresis-free capacitance enhancement in a ferroelectric-dielectric combination. However, when a semiconductor is placed in series with the ferroelectric, due to strong mismatch of charge between the two, it becomes difficult to obtain such gain avoiding hysteresis, particularly in the subthreshold regime, and hence achieving a subthreshold slope below 60 mV/decade becomes difficult to achieve in hysteresis free manner. However, such a structure may be useful to reduce the overdrive voltage of the transistor, eventually reducing the chip supply voltage. Searching for ultra-thin, low polarization ferroelectric with a large band offset with the semiconductor will have important technological implications in this direction.


\begin{thebibliography}{10}
\bibitem{tnt10}
T. N. Theis and P. M. Solomon, \lq\lq{}In quest of the \lq{}next switch\rq{}: Prospects
for greatly reduced power dissipation in a successor to the silicon field effect
transistor,\rq\rq{} Proc. IEEE, 98, 2005–2014, 2010.
\bibitem{ion11}
A. M. Ionescu	and H. Riel, \lq\lq{}Tunnel field-effect transistors as energy-efficient electronic switches,\rq\rq{} 479, 329-337, 2011.
\bibitem{gopal02}
K. Gopalakrishnan, P. B. Griffin, and J. D. Plummer, \lq\lq{}I-MOS: A novel
semiconductor device with a subthreshold slope lower than kT/q,\rq\rq{} in
Proc. IEEE IEDM, Dec. 2002, pp. 289-292.
\bibitem{na05}
N. Abele, R. Fritschi, K. Boucart, F. Casset, P. Ancey, and A. M. Ionescu,
\lq\lq{}Suspended-gate MOSFET: Bringing new MEMS functionality into
solid-state MOS transistor,\rq\rq{} in Proc. IEEE IEDM, Dec. 2005,
pp. 479-481.
\bibitem{Kam05}
H. Kam, D. T. Lee, R. T. Howe, and T.-J. King, \lq\lq{}A new nano-electromechanical
field effect transistor (NEMFET) design for low-power
electronics,\rq\rq{} in Proc. IEEE IEDM, Dec. 2005, pp. 463-466.
\bibitem{ss08}
S. Salahuddin and S. Datta, \lq\lq{}Use of Negative Capacitance to Provide Voltage Amplification for Low Power Nanoscale Devices,\rq\rq{} Nano Lett., 8, 405-410, 2008.
\bibitem{dj10}
D. Jimenez, E. Miranda, and A. Godoy, \lq\lq{}Analytic model for the
surface potential and drain current in negative capacitance fieldeffect
transistors,\rq\rq{} IEEE Trans. Elec. Dev., 57, 2405-2409, 2010.
\bibitem{chen11}
H.-P. Chen, V. C. Lee, A. Ohoka, J. Xiang, and Y. Taur, \lq\lq{}Modeling
and design of ferroelectric MOSFETs,\rq\rq{} IEEE Trans. Elec. Dev.,
58, 2401-2405, 2011.
\bibitem{khan11iedm}
A. I. Khan, C. W. Yeung, C. Hu, and S. Salahuddin, \lq\lq{}Ferroelectric
negative capacitance MOSFET: Capacitance tuning and antiferroelectric
operation,\rq\rq{} in Proc. IEDM, 2011, pp. 255-258.
\bibitem{frank14}
D. J. Frank, P. M. Solomon, C. Dubourdieu, M. M. Frank, V. Narayanan, and T. N. Theis, \lq\lq{}The Quantum Metal Ferroelectric Field-Effect Transistor,\rq\rq{} IEEE Trans. Elec. Dev., 61, 2145-2153, 2014.
\bibitem{jain14}
A. Jain and M. A. Alam, \lq\lq{}Stability Constraints Define the Minimum Subthreshold Swing of a Negative Capacitance Field-Effect Transistor,\rq\rq{} IEEE Trans. Elec. Dev., 61, 2235-2242, 2014.
\bibitem{khan11apl}
A. I. Khan, D. Bhowmik, P. Yu, S. J. Kim, X. Pan, R. Ramesh, and S. Salahuddin, \lq\lq{}Experimental evidence of ferroelectric negative capacitance in nanoscale heterostructures,\rq\rq{} Appl. Phys. Lett., 99, 113501, 2011.
\bibitem{appleby14}
D. J. R. Appleby, N. K. Ponon, K. S. K. Kwa, B. Zou, P. K. Petrov, T. Wang, N. M. Alford, and A. ONeill, \lq\lq{}Experimental Observation of Negative Capacitance in Ferroelectrics
at Room Temperature,\rq\rq{} Nano Letters, 14, 3864-3868, 2014.
\bibitem{gas08}
G. A. Salvatore, D. Bouvet, and A. M. Ionescu, \lq\lq{}Demonstration of subthreshold swing smaller than 60 mV/decade in Fe-FET with P(VDF- TrFE)/SiO2 gate stack,\rq\rq{} in Proc. IEEE IEDM , Dec. 2008, pp. 167-170.
\bibitem{rusu10}
A. Rusu, G. A. Salvatore, D. Jiménez, et al., \lq\lq{}Metal-ferroelectric- metal-oxide-semiconductor field effect transistor with sub-60 mV/decade subthreshold swing and internal voltage amplification,\rq\rq{} in IEDM Tech. Dig., 2010, pp. 395-398.
\bibitem{sdUn}
S. DasGupta, A. Rajashekhar, K. Majumdar, N. Agrawal, A. Razavieh, S. Trolier-McKinstry, and S. Datta, \lq\lq{}Sub-kT/q Switching in Strong Inversion in PbZr$_{0.52}$Ti$_{0.48}$O$_3$ Gated Negative Capacitance FETs,\rq\rq{} IEEE Journal on Exploratory Solid-State Computational Devices and Circuits, Vol. 1, pp. 43-48, 2015.
\bibitem{lgbook}
M. E. Lines and A. M. Glass, \lq\lq{}Principles and applications of ferroelectrics and related materials,\rq\rq{} Oxford University Press, 2001.
\bibitem{rabebook}
K. M. Rabe, C. H. Ahn and J. M. Triscone, \lq\lq{}Physics of ferroelectrics: a modern perspective,\rq\rq{} Springer, 2007.
\bibitem{wangJap07}
Y. L. Wang, A. K. Tagantsev, D. Damjanovic, and N. Setter, V. K. Yarmarkin, A. I. Sokolov, I. A. Lukyanchuk, \lq\lq{}Landau thermodynamic potential for BaTiO$_3$,\rq\rq{} J. Appl. Phys., 101, 104115, 2007.
\bibitem{taurbook}
Y. Taur and T. H. Ning, \lq\lq{}Fundamentals of modern VLSI devices,\rq\rq{} Cambridge Univ. Press, 1998.
\bibitem{d2000}
S. Ducharme, V. M. Fridkin, A.V. Bune, S. P. Palto, L. M. Blinov, N. N. Petukhova, and S. G. Yudin, \lq\lq{}Intrinsic Ferroelectric Coercive Field,\rq\rq{} Phys. Rev. Lett., 84, 175-178, 2000.
\end{thebibliography}
\end{document}